\documentclass[reprint, aps, pra, superscriptaddress, longbibliography, nofootinbib]{revtex4-1}
\usepackage{epstopdf}
\usepackage{graphicx}
\usepackage{amsfonts}
\usepackage{amssymb}
\usepackage{amsthm}
\usepackage{array}
\usepackage{amsmath}
\usepackage{verbatim} 
\usepackage[colorlinks, linkcolor=blue, anchorcolor=blue, citecolor=blue, urlcolor=blue]{hyperref}
\usepackage{color}
\usepackage{bbold}
\usepackage{mathtools}
\usepackage[normalem]{ulem}

\newtheorem{observation}{Observation}

\newcommand{\ket}[1]{\left\vert#1\right\rangle}
\newcommand{\bra}[1]{\left\langle#1\right\vert}

\def\bra#1{\langle #1|}
\def\ket#1{\left|#1 \right>}

\def\Tr{\mbox{Tr}}

\begin{document}
\title{Engineering a heat engine purely driven by quantum coherence}
\author{Stefan Aimet}
\email{stefan.aimet@gmail.com}
\affiliation{Quantum Optics and Laser Science, Blackett Laboratory, Imperial College London, London SW7 2AZ, United Kingdom}
\author{Hyukjoon Kwon}
\email{hjkwon@kias.re.kr}
\affiliation{School of Computational Sciences, Korea Institute for Advanced Study, Seoul 02455, South Korea}

\begin{abstract}
The question of whether quantum coherence is a resource beneficial or detrimental to the performance of quantum heat engines has been thoroughly studied but remains undecided. To isolate the contribution of coherence, we analyze the performance of a purely coherence-driven quantum heat engine, a device that does not include any heat flow during the thermodynamic cycle. The engine is powered by the coherence of a multiqubit system, where each qubit is charged via interaction with a coherence bath using the Jaynes-Cummings model. We demonstrate that optimal coherence charging and hence extractable work is achieved when the coherence bath has an intermediate degree of coherence. In our model, the extractable work is maximized when four copies of the charged qubits are used. Meanwhile, the efficiency of the engine, given by the extractable work per input coherence flow, is optimized by avoiding the coherence being stored in the system-bath correlations that is inaccessible to work. We numerically find that the highest efficiency is obtained for slightly lower temperatures and weaker system-bath coupling than those for optimal coherence charging.

\end{abstract}
\maketitle

\section{Introduction}
\label{section:Intro}
The advent of nanoscale technologies has made it increasingly important to understand and potentially exploit quantum phenomena, such as quantum entanglement and coherence~\cite{Jaeger2018, Deutsch2020, Dowling2003}. This has raised the question of whether thermodynamic laws, which describe physics at the macroscopic level, will still be valid when quantum mechanical principles at the microscopic level become significant~\cite{gemmer2004quantum, Binder2018}. One of the main goals in the field of quantum thermodynamics is to identify the role that quantum effects play in thermodynamics, which will help us better understand devising technologies in the quantum regime.

Quantum coherences with respect to the system's energy eigenstates constitute a resource~\cite{Streltsov2017} in the resource-theoretic approach of quantum thermodynamics. In this framework, a family of second laws has been identified via generalized free energies that impose constraints on the allowed state transformations~\cite{Horodecki2013, PhysRevLett.115.210403, Brandao2015}. Such an approach has been further extended to study nonequilibrium thermodynamics of quantum processes ~\cite{PhysRevE.92.032129, PhysRevX.6.041017, PhysRevX.8.011019, Kwon2019}. In particular, quantum coherence has been thoroughly investigated in the context of work extraction~\cite{Scully2003, Horodecki2013, Skrzypczyk2013, Lostaglio2015DescriptionOQ, PhysRevX.5.041011, Kwon2018, Klatzow2019}. One of the first steps towards exploiting quantum coherence for work extraction was made in Ref.~\cite{Scully2003} by coupling qutrits to a single bath charged with coherence. However, the results of this paper are still in accordance with the classical Carnot efficiency requiring a temperature gradient between the system and bath for the efficiency to be nonzero. Quantum coherence was also shown to enhance the charging process of batteries~\cite{GarciaPintos2020, Culhane2022}. More recently, the coherent contribution of \textit{ergotropy}, which is the maximum extractable work yield over a cyclical variation of the parameters of the system Hamiltonian, was examined in Refs.~\cite{Francica2020,Lobejko2022}.

Most existing proposals for quantum heat machines couple the system to baths at two different temperatures, causing heat flow~\cite{Goswami2013, Uzdin2015, Latune2020, Santos2021}. In some cases, it was also shown that coherence can be detrimental to the efficiency~\cite{Brandner2017,Hammam2021} or does not give any advantage over classical resources in terms of maximum extractable work~\cite{Strasberg2017}. Therefore, the question remains whether quantum coherence is beneficial to thermodynamic processes in a universal way or is limited to specific instances~\cite{Binder2018, Latune1_2019, Latune2_2019}.

In this paper, we go a step further to elucidate the role of quantum coherence in work extraction by examining a heat engine purely driven by quantum coherence.

The performance of such a coherence-driven engine does not rely on the concept of heat flow. Instead, we devise an engine system that operates at the same temperature as a coherence bath throughout the protocol, in the sense that classical energy distributions are invariant under the interaction. As shown hereby, quantum coherence cannot only enhance work extraction for a thermodynamic cycle but can fully sustain it without changing the energy distribution of the system. To this end, we design a protocol for a multiqubit quantum coherence-driven heat engine. There is no classical analog for such an engine operating at the same temperature as the bath. By adopting the Jaynes-Cummings model to describe the interaction between each qubit and the bosonic coherence bath, we investigate the performance of the charging mechanism, with coherence being accessible for work extraction. It turns out that optimal charging and optimal efficiency occur for a small number of qubits at intermediate temperatures.

The paper is organized as follows. In Sec.~\ref{section:Coherence-driven heat engine cycle}, basic notions for work extraction from a quantum coherence-driven engine are introduced alongside a description of running the thermodynamic cycle. Section~\ref{section:Coherence charging of individual qubits} investigates the coherence charging mechanism of individual qubits. Finally, Sec.~\ref{section:Thermodynamic analysis of the coherence-driven heat engine} gives a thermodynamic analysis of the device.

\section{Coherence-driven heat engine cycle}
\label{section:Coherence-driven heat engine cycle}
\subsection{Relationship between coherence and work extraction}
\subsubsection{Coherence}
The classical thermodynamic properties of a quantum state $\rho$ with Hamiltonian $H$ and its energy distribution are contained in the diagonal elements with respect to the energy eigenstates. For this purpose, let us consider the \textit{fully dephased state}~\cite{Lostaglio2015, Horodecki2013, Kwon2018} corresponding to $\rho$ defined by removing all of its off-diagonal elements:
\begin{equation}
\Delta(\rho):=\sum_{E_i}\ket{E_i}\bra{E_i}\bra{E_i}\rho\ket{E_i},
\end{equation}
where $\ket{E_i}$ are the energy eigenstates of the system. However, the fact that a quantum state can be in a superposition between eigenstates leads to the concept of coherence as a resource describing the off-diagonal elements of the density matrix.

Meanwhile, the \textit{dephasing} operation defined as
\begin{equation}
\mathcal{D}(\rho):=\sum_E P_E\rho P_E
\end{equation}
projects out only the energy block-diagonal part of $\rho$ using the projectors onto each distinct energy subspace $P_E:=\sum_{H \ket{E_\mu} = E\ket{E_\mu}} \ket{E_\mu}\bra{E_\mu}$. This state is only left with \textit{internal coherence}~\cite{Kwon2018,Mendes2019} of a state $\rho$ which quantifies the coherences between states of equal energies that are unaffected by the action of $\mathcal{D}$. The internal coherence can be quantified by the relative entropy between the dephased state  $\mathcal{D}(\rho)$ and the fully dephased state $\Delta(\rho)$ as
\begin{equation}
   C_{\rm int} = S(\mathcal{D}(\rho) \| \Delta(\rho)) = S(\Delta(\rho))-S(\mathcal{D}(\rho)),
   \label{eq:Coh_int}
\end{equation}
where $S(\rho \| \sigma) = \Tr[\rho(\ln\rho - \ln\sigma)]$ is the quantum relative entropy and $S(\rho)=-\Tr[\rho\ln{\rho}]$ is the von Neumann entropy. The action of the dephasing operation $\mathcal{D}$ removes \textit{external coherence} $C_{\rm ext}(\rho)$ which measures coherence between states of different energy~\cite{Lostaglio2015DescriptionOQ, Mendes2019, Kwon2018}:
\begin{equation}
C_{\rm ext}(\rho) = S(\mathcal{D}(\rho)) - S(\rho).
\label{eq:extcoh}
\end{equation}
We note that the total coherence, quantified by the relative entropy of coherence, is obtained by adding the two different types of coherences, i.e., $C_{\rm tot}(\rho)= S(\rho \| \Delta(\rho))=C_{\rm int}(\rho)+C_{\rm ext}(\rho)$~\cite{Lostaglio2015DescriptionOQ}.

\subsubsection{Work extraction}
Suppose that a quantum system $\rho_S$ with Hamiltonian $H_S$ is in contact with a thermal bath with inverse temperature $\beta = (k_B T)^{-1}$. At thermal equilibrium, the system state is in the Gibbs state $\gamma_S = e^{-\beta H_S}/Z_S$ with $Z_S = \Tr [e^{-\beta H_S}]$. When a nonequilibrium quantum state $\rho_S$ relaxes to the equilibrium state $\gamma_S$, work can be extracted by interacting with the bath. In classical thermodynamics, work is a result of heat transfer between the system and the bath, which is entirely captured by the dynamics of the incoherent, fully dephased state $\Delta(\rho_S)$. In quantum theory, however, it has been repeatedly shown that genuine quantum features like entanglement or coherence can have an effect on work extraction~\cite{Korzekwa2016,hovhannisyan2013entanglement}.

The incoherent work contribution is the work extractable which only stems from a change of the diagonal elements of a quantum state. It is sourced by a classical heat flow~\cite{Kwon2018} changing the classical energy distribution as given by the fully dephased state $\Delta(\rho_S)$. On the other hand, coherence can also contribute to work, which has no classical counterpart and only arises due to a change of the off-diagonal elements of $\rho_S$. As outlined earlier, there are two different types of coherence, internal and external coherence, that may contribute to this coherent work extraction.

It was studied~\cite{Lostaglio2015DescriptionOQ, Kwon2018} that only internal coherence can be utilized for coherent work extraction. Having charged each of the $N$ copies of $\rho_{S}$ with coherence, the dephasing operation is applied to the collective state $\rho_{S}^{\otimes N}\rightarrow \mathcal{D}(\rho_{S}^{\otimes N})$, with the lost (external) coherence being locked and inaccessible to work extraction. The work extracted from internal coherence can then be stored in a battery-like weight. The average amount of work~\cite{Skrzypczyk2013} that can be extracted from coherence for a system at inverse temperature $\beta$ is then shown to be directly proportional to its internal coherence
\begin{equation}
    W_{\rm coh}(\rho^{(N)}_S)=\frac{1}{\beta}C_{\rm int}(\rho^{(N)}_S),
    \label{eq:coherentwork}
\end{equation}
where $\rho^{(N)}_S$ denotes the density operator of the system state for $N$ qubits. In particular, for a quantum system, $\rho_S$ with a nondegenerate Hamiltonian coherence cannot be converted into work as all the coherence stored in the state becomes external coherence, an effect known as \textit{work-locking}. Nevertheless, internal coherence can be \textit{activated} by putting multiple copies of quantum states together~\cite{Lostaglio2015DescriptionOQ}.

While earlier work~\cite{Tajima2021, Shi_2020, Monsel2020} included both heat flow and coherence flow as a source of work, we will restrict ourselves to cases that involve solely the latter. This is relevant as it demonstrates, also in an experimental setting, that work can be entirely sourced by coherence as a fundamental resource. We focus on the case, where the quantum state $\rho_S^{(N)}$ has the same energy distribution as the Gibbs state, i.e., $\Delta(\rho^{(N)}_S)=\gamma^{\otimes N}_S$. In this case, classical heat flow and incoherent work generation are excluded, but only coherent work $W_{\rm coh}$ contributes to the total extractable work.

\begin{figure}
    \centering
    \includegraphics[width=.9\linewidth]{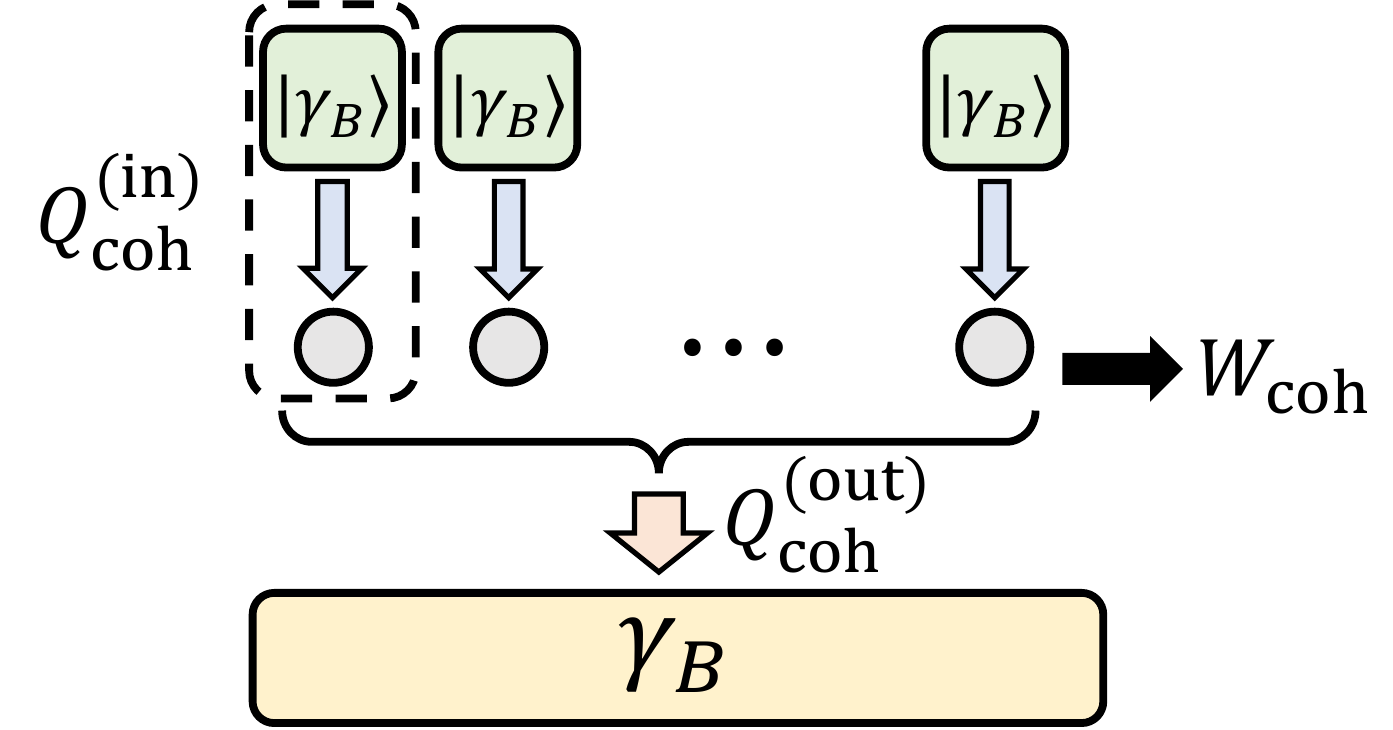}
    \caption{Schematic showing a coherence-driven heat engine protocol using the sequential coherence charging of individual qubits at uniform inverse temperature $\beta$. Each qubit is charged sequentially to a state $\rho_S$ with coherence by input coherence flow $Q_{\rm coh}^{\rm (in)}=\sum_{k}Q^{{\rm (in)}}_{{\rm coh},k}$, where $Q^{{\rm (in)}}_{{\rm coh}, k} = - k_B T \Delta C_{\rm ext} (\rho_B)$ corresponds to the change of coherence of the coherence bath $\ket{\gamma_B}$ used for the $k$th qubit weighted by its temperature (see the main text for details). The coherence flow is partially stored in the total system via internal coherence and external coherence but is also stored in the system-bath correlations. While only the internal coherence $C_{\rm int}(\rho_S^{\otimes N})$ can be extracted as coherent work $W_{\rm coh} = k_B T C_{\rm int}(\rho_S^{\otimes N})$, the external coherence $C_{\rm ext}(\rho_S^{\otimes N})$ is locked and erased from the system during work extraction while the coherence stored in the system-bath correlations is allowed to flow as $Q_{\rm coh}^{\rm (out)}$ into a thermal bath $\gamma_B$.}
    \label{fig:Concept}
\end{figure}

\subsection{Basics for engineering the protocol}
We focus on a noninteracting $N$-qubit chain, where each qubit has two energy levels of $\pm \omega_0/2$ with the resonance frequency $\omega_0$. In this paper, we work in units with $\hbar=1$. The total system Hamiltonian is $H_S^{(N)} = (\omega_0/2)\sum_{k=1}^{N}\sigma_{z|k}$ with the Pauli operator $\sigma_{z|k}$ acting on the $k$th qubit. We initialize each qubit of the system to the Gibbs state  $\gamma_S=\left(\ket{0}\bra{0}+e^{-\beta\omega_0}\ket{1}\bra{1}\right)/Z_S$ with partition function $Z_S = 1 + e^{-\beta\omega_0}$. Thus, the total system state becomes $\rho^{(N)}_{S}(0) = \gamma_S^{\otimes N}$ at time $t=0$.

We design a coherence charging protocol that does not induce energy flow from the bath to the system. This can be done by coupling the system to a coherence bath with the same temperature in the quantum state
$$
\ket{\gamma_B} = Z_B^{-1/2} \sum_i e^{-\beta \epsilon_i /2} \ket{\epsilon_i},
$$
where the Hamiltonian of the bath $H_B = \sum_i \epsilon_i \ket{\epsilon_i}\bra{\epsilon_i}$ is assumed non-degenerate and $Z_B = {\rm Tr}[e^{-\beta H_B}] = \sum_i e^{-\beta \epsilon_i}$. We call $\ket{\gamma_B}$ a coherent bath state as the state has the same energy distribution as the Gibbs state, i.e., $\Delta(\ket{\gamma_B}\bra{\gamma_B}) = \gamma_B = e^{-\beta H_B}/Z_B$.

In general, we take the interaction to have the following form:
\begin{equation} \label{eq:H_t}
    H(t) = H^{(N)}_S + H_B + H^{(N)}_I(t),
\end{equation}
where $[H(t), H^{(N)}_S + H_B] = 0$ for all $t$ with interaction Hamiltonian $H^{(N)}_I(t)$ so that the total energy of the system and the bath is conserved throughout the system-bath interaction. The resulting state after time $t$ then becomes
$$
\rho_{SB}(t) = U(t) \left( \gamma^{\otimes N}_S \otimes \ket{\gamma_B}\bra{\gamma_B} \right) U^\dagger(t),
$$
where $U(t) = {\cal T} \left[ e^{-i \int_0^t H(t') dt' }\right]$ is the time evolution operator with time-ordering ${\cal T}$. 

Consequently, the local states of the system and bath after time $t$ become $\rho^{(N)}_S(t) = {\rm Tr}_B \left[ \rho^{(N)}_{SB}(t)\right]$ and $\rho_B(t) = {\rm Tr}_S \left[ \rho^{(N)}_{SB}(t)\right]$, respectively. In this case, we note that energy distributions of both system and bath states do not change during the interaction, which can be generalized to the following observation: 
\begin{observation} When an initial Gibbs state $\gamma^{\otimes N}_S$ is interacting with a single coherence bath $\rho_B$ satisfying $\Delta(\rho_B) = \gamma_B$ under the energy preserving interaction Hamiltonian $H_I$, the diagonal elements of the system after time $t$ are unchanged.
\label{ob:observation1}
\end{observation}
The technical proofs throughout the paper can be found in the Appendices. While the energy distributions on the diagonal remain the same, off-diagonal elements become nonzero because of the interaction for $t>0$. This can be interpreted as coherence stored in the coherent bath state being transferred to the system, thus charging the system with coherence. We also note that Observation~\ref{ob:observation1} strongly depends on the initial conditions for the system state $\rho_S^{(N)}(0) = \gamma_S^{\otimes N}$. If the initial state has diagonal elements that are not the same as the Gibbs state ($\Delta(\rho_S^{(N)}(0)) \neq \gamma_S^{\otimes N}$) or contains nonzero off-diagonal elements ($\Delta(\rho_S^{(N)}(0)) = \gamma_S^{\otimes N} \neq \rho_S^{(N)}(0)$), the system's diagonal elements after the interaction may not remain the same as those for the initial state, i.e., $\Delta (\rho_S^{(N)}(t)) \neq \rho_S(0)$ or $\Delta (\rho_S^{(N)}(t)) \neq \gamma_S^{\otimes N}$ (see Appendix~\ref{appendix:a} for more details).

As an illustrative example, let us consider the internal coherence of the maximally coherent state for $N=2$ qubits. The dephased, or block-diagonalized, state at time $t$ becomes
\begin{equation}
{\cal D}(\rho_S) = 
\frac{e^{-\beta\omega_0}}{Z_S^2}
\left(
\begin{matrix}
 e^{-\beta \omega_0} & 0 & 0 & 0  \\
0 & 1& \alpha^*(t) & 0 \\
0 & \alpha(t) & 1 & 0 \\
0 & 0 & 0 &  e^{\beta \omega_0}
\end{matrix}
\right)
\end{equation}
with some factor $0 \leq |\alpha(t)| \leq 1$. The case of $|\alpha|=0$ corresponds to the thermal state providing zero work and $|\alpha|=1$ to the fully (internally) coherent state providing maximal work. For a given value of $\beta$, the upper bound of internal coherence that can be harnessed for coherent work extraction can be calculated from $N$ copies of the maximally coherent state $\ket{\gamma_S}^{\otimes N}$, where $\ket{\gamma_S} = (\ket{0} + e^{-\beta\omega_0/2} \ket{1})/\sqrt{Z_S}$.

\section{Coherence charging of individual qubits}
\label{section:Coherence charging of individual qubits}
\subsection{Jaynes-Cummings model}
To illustrate the main characteristics of coherence charging, we consider the simplest model, in which each qubit is charged sequentially. A widely studied example for the interaction Hamiltonian $H_I$ in Eq.~\eqref{eq:H_t}, now specified for a single qubit, is used in the Jaynes Cummings model~\cite{Jaynes1963}. It describes the interaction between a two-level system described by $H_S = (\omega_0/2) \sigma_z$ and an infinite dimensional bosonic system with Hamiltonian $H_B = \omega_0 a^\dagger a$, where $a^{\dagger}$ and $a$ are the bosonic creation and annihilation operators. Each qubit interacts via
\begin{equation}
    H_I = g\left(\sigma_{-}a^{\dagger}+\sigma_{+}a\right),
    \label{eq:JC_interaction}
\end{equation}
where $g$ is the coupling strength and $\sigma_{+}=\ket{1}\bra{0}$ and $\sigma_{-}=\ket{0}\bra{1}$ are the fermionic creation and annihilation operators of the system, respectively. The coherent bath state for a bosonic system is given as $\ket{\gamma_B} = Z_B^{-1/2} \sum_{n=0}^{\infty} e^{-\beta \omega_0 n /2} \ket{\epsilon_n}$ with $Z_{B}=(1-e^{-\beta\omega_0})^{-1}$.

After charging one qubit, the coherence bath is repumped, and the process repeats for the remaining qubits (see Fig.~\ref{fig:Concept}). Given a constant coupling strength $g$, the total time evolution operator $U(t) = U_I(t) U_0(t)$ for each qubit is a product of the Schr\"{o}dinger picture unitary $U_0(t)=e^{-i (H_S + H_B) t}$ and the interaction picture unitary $U_I(t)=e^{-iH_I t}$. An explicit form of the time evolution operator in the interaction picture is given by~\cite{Klimov2009}:
\begin{equation}
\begin{aligned}
U_I(t) = \cos{\left(gt\sqrt{\hat{n}+1}\right)}\ket{1}\bra{1} + \cos{\left(gt\sqrt{\hat{n}}\right)}\ket{0}\bra{0}\\ - i\frac{\sin{\left(gt\sqrt{\hat{n}+1}\right)}}{\sqrt{\hat{n}+1}}a\ket{1}\bra{0} - i\frac{\sin{\left(gt\sqrt{\hat{n}}\right)}}{\sqrt{\hat{n}}}a^{\dagger} \ket{0}\bra{1},
\label{eq:UI}
\end{aligned}
\end{equation}
where $\hat{n} = a^\dagger a$ is the number operator of the bosonic bath. Evolving the initial total state for each qubit and the bath $\rho(0)=\gamma_S\otimes\rho_B$ with time $t$, where $\rho_B=\ket{\gamma_B}\bra{\gamma_B}$, and tracing out the bath degrees of freedom, one arrives at the system density operator
\begin{equation}
\rho_S(t)=\gamma_S+\frac{1}{Z_S}\left(e^{i\omega_0 t}\delta^*(t)\ket{0}\bra{1}+e^{-i\omega_0 t}\delta(t)\ket{1}\bra{0}\right),
\label{eq:rho after interaction}
\end{equation}
where
\begin{equation}
\begin{aligned}
    \delta(t)=\frac{i}{Z_B}\sum_{p=0}^{\infty} e^{-\beta\omega_0 p} \sin{\left(gt(\sqrt{p+1}-\sqrt{p})\right)}.
\end{aligned}
\label{eq:single_rho01}
\end{equation}
We note that $0\leq |\delta(t)|\leq 1$. If we charge each qubit individually, we only charge external coherences, as there is no degeneracy in a single-qubit Hamiltonian $H_S$. Internal coherences can then be activated by putting multiple qubits together.

\begin{figure*}
\includegraphics[scale=0.58]{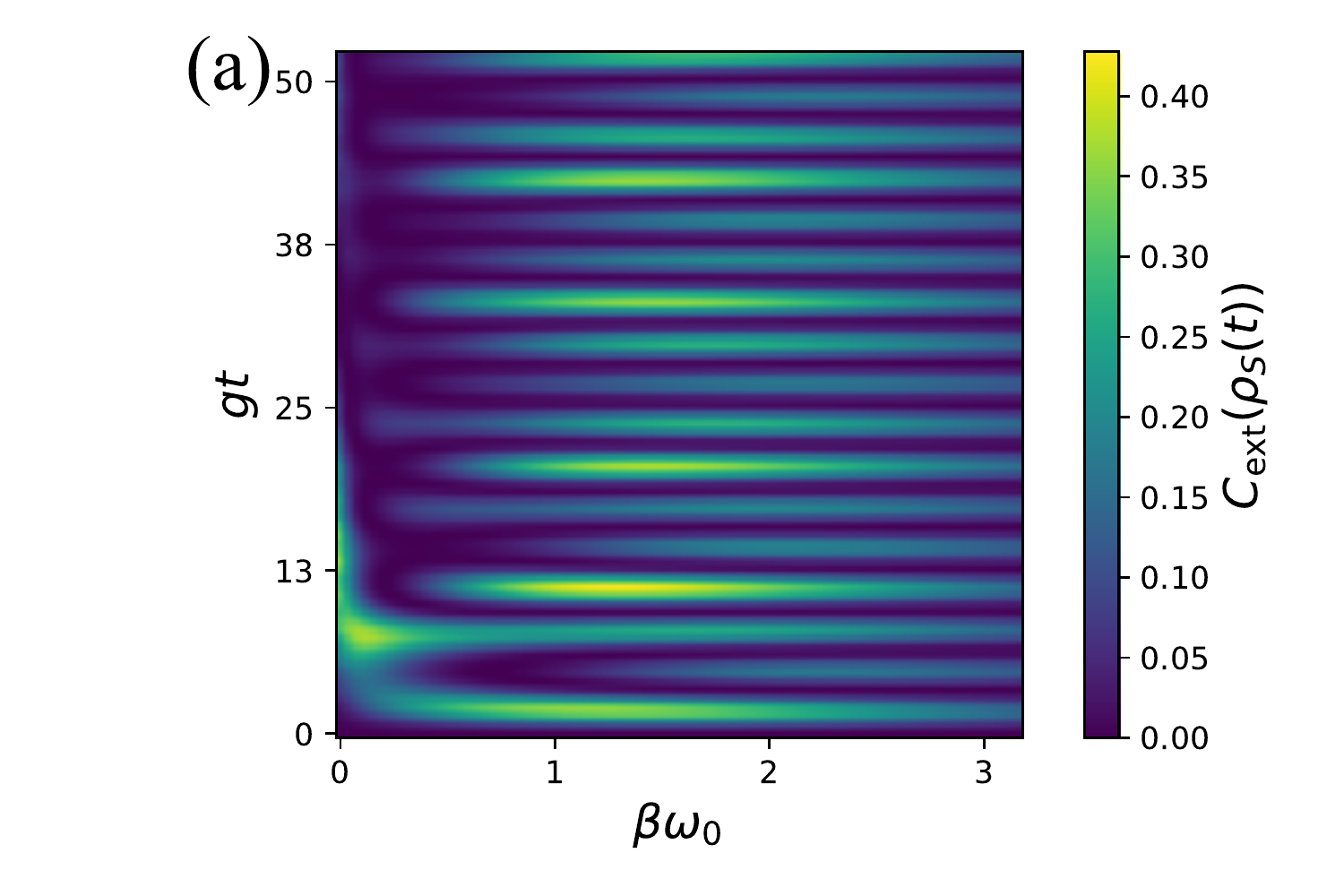}
\includegraphics[scale=0.58]{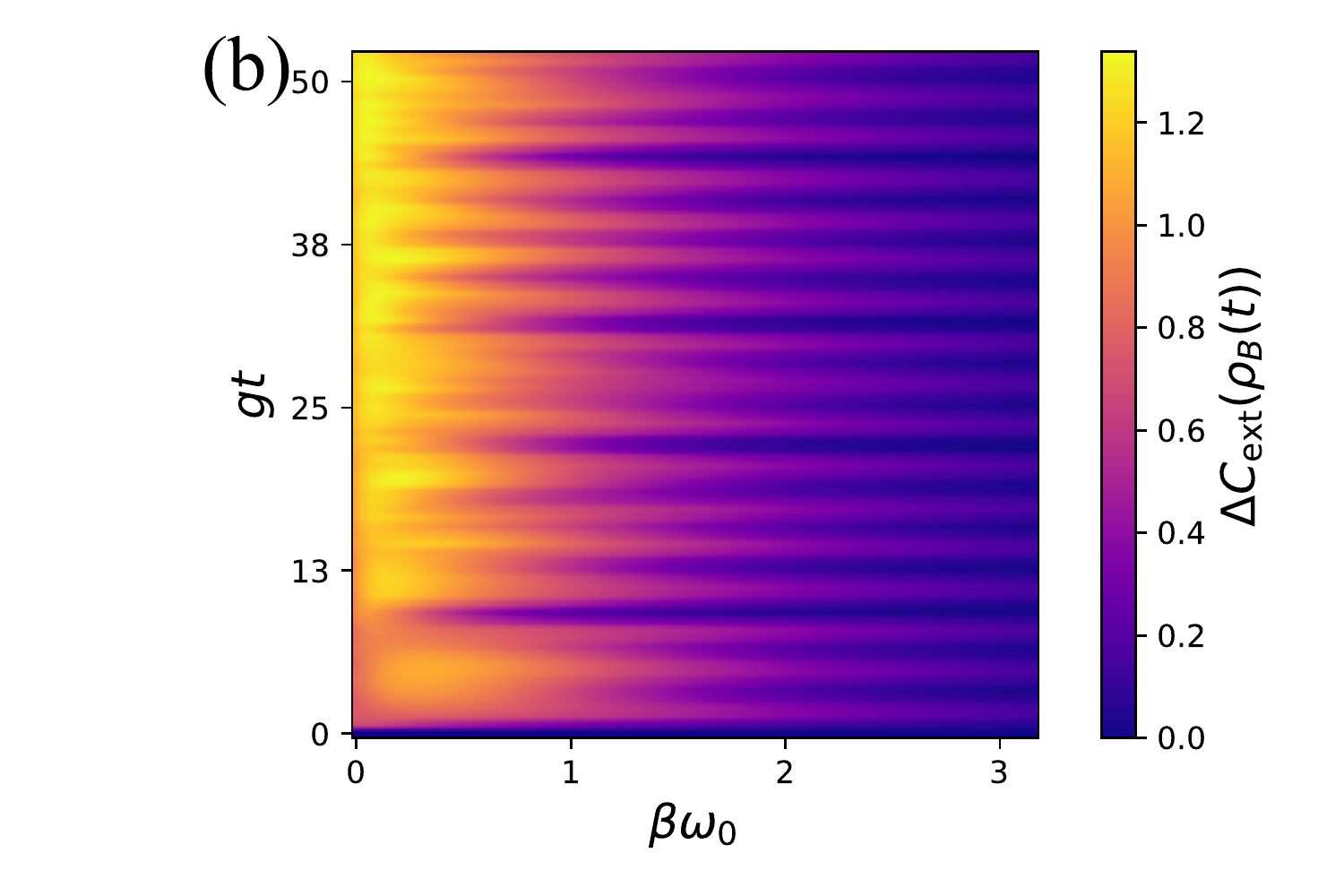}
\caption{The coherence of (a) the system $C_{\rm ext}(\rho_S(t))$ and (b) the coherence change of the bath $\Delta C_{\rm ext}(\rho_B(t))$ as a function of inverse temperature $\beta\omega_0$ and coupling strength $gt$ for individual qubit charging via the Jaynes-Cummings interaction. As both the system and bath have non-degenerate energy levels, only the external coherence contributes to the total amount of coherence after the interaction. Higher temperatures increase the coherence change of the bath, which we can interpret as coherence being locked in the system-bath correlations (see Eq.~\eqref{eq:coh_conserv}).}
\label{fig:Coh_charge}
\end{figure*}

One may ask whether we can devise a more efficient charging protocol by charging internal coherence explicitly already during the interaction and not only activating it \textit{a posteriori}. This can avoid charging external coherence inaccessible to work extraction. However, we justify the use of coherence charging for individual qubits at constant coupling strength as follows. First, we stress that no advantage can be demonstrated by assuming a collective Tavis-Cummings interaction (TC)~\cite{Tavis1968} for a single coherent bath. The TC interaction Hamiltonian is of the form $H^{(N)}_{I}=\sum_{k=1}^{N}g_k\sigma_{+|k}a+g^{*}_k\sigma_{-|k}a^{\dagger}$, where the coupling strengths $g_k\in\mathbb{C}$ can be tuned. The fact that the individual qubit charging always outperforms the TC interaction relies upon the repumping of the bath after each qubit interaction and discarding the system-bath correlations. We formalize this in the following observation:
\begin{observation}
\label{ob:observation2}
If multiple copies of the Gibbs state $\gamma_S^{\otimes N}$ are interacting with a single coherence bath $\rho_B$ satisfying $\Delta(\rho_B) = \gamma_B$ under the energy-preserving interaction Hamiltonian $H^{(N)}_I$, only the external coherence of the system can be charged.
\end{observation}

Secondly, the coherence charging of a two-level system cannot be improved by admitting time-dependent coupling strengths $g(t)$ and any numerical optimal control techniques. This follows as the unitary evolution operator $U(t)=\mathcal{T}\left[ e^{-i\int_{0}^{t}H(t')dt'}\right] = e^{- i \chi (\sigma_- a^\dagger + \sigma_+ a)} e^{-i (H_S + H_B)t}$ can always be expressed as a product of evolution operators for time-independent coupling strengths with $\chi = \int_0^t g(t') dt'$ as $[H_S + H_B, H_I]=0$ for every $g(t)$.

Let us now consider how to optimize the coherence charging protocol for each qubit at constant coupling strength. If the external coherence $C_{\rm ext}(\rho_S(t))$ is maximized in our charging protocol, then accordingly also the internal coherence of a chain of qubits is maximized. Investigating the temperature dependence of the coherence elements $|\delta(t)|/Z_S$ being charged, we see that no coherence is charged for both the zero and infinite temperature limit, i.e., $\lim_{\beta \to 0,\infty}|\delta(t)|/Z_S$=0. Notably, this is in sharp contrast to the conclusion drawn from the maximally coherent state $\ket{\gamma_S}\bra{\gamma_S}^{\otimes N}$, where internal coherence is maximized at infinite temperature. Instead, the Jaynes-Cummings model exhibits maximum coherence charging for an intermediate temperature.

\subsection{Effective finite-size bath approximation}
To numerically investigate the dependence of $|\delta(t)|/Z_S$ as a measure for external coherence, we invoke the \textit{effective bath approximation}, which allows us to approximate the infinite-dimensional bath by a finite-size bath of \textit{effective dimension} $d^*$ (see Appendix~\ref{appendix:c} for further details). While baths are often assumed to be infinite-dimensional reservoirs, practical considerations force us to compute expressions such as in Eq.~\eqref{eq:single_rho01} numerically with some finite precision when no analytically exact answer can be found. The exponential in the sum in Eq.~\eqref{eq:single_rho01} strongly suppresses the significance of higher-order terms for low temperatures. Physically, this corresponds to the freeze-out of higher-lying states, and thus this provides a fully accurate description of the low-temperature regime. In contrast, higher temperatures also couple the system to higher-lying states of the bath. This intuitive explanation motivates the conclusion found for the Jaynes-Cummings model where no coherence is generated for the two extreme limits of zero and infinite temperature regimes. In general, the effective bath dimension $d^*$ is an increasing function of coupling strength $gt$ and a decreasing function of temperature $\beta\omega_0$.

As shown in Fig.~\ref{fig:Coh_charge}(a), the external coherence $C_{\rm ext}(\rho_S(t))$ is a non-trivial function of the parameter pair $(\beta\omega_0, gt)$. Qualitatively, the dynamics of external coherence show recurrent behavior. The corresponding optimization is also non-trivial, as one may find high coherence charging for a very large coupling strength. Due to the increase in system-bath correlations with coupling strength as shown in Sec.~\ref{section:Thermodynamic analysis of the coherence-driven heat engine}, only weak coupling is considered for the optimization of efficiency later on.

\begin{figure}
    \centering
    \includegraphics[scale=0.5]{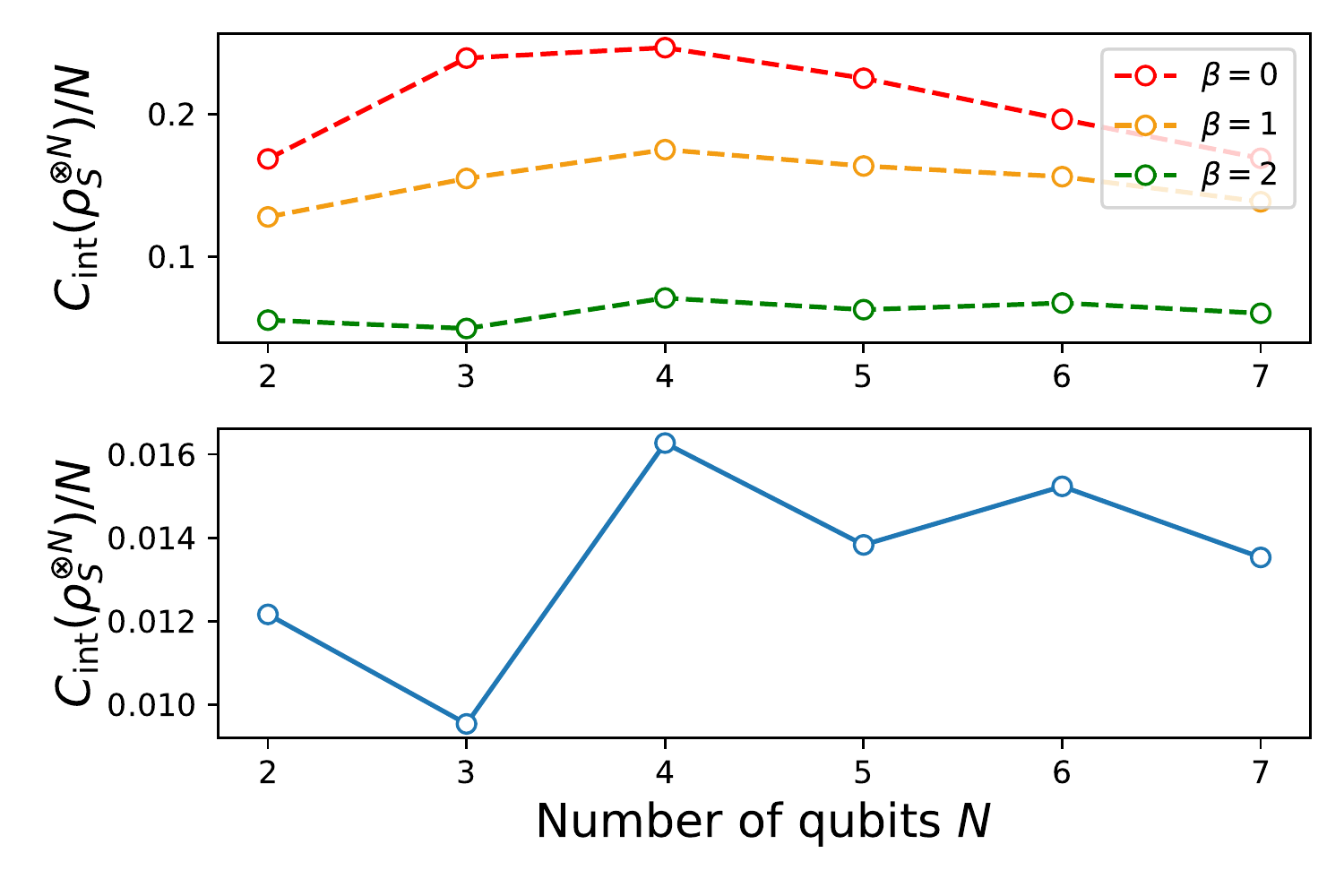}
    \caption{The internal coherence per qubit $C_{\rm int}(\rho_S^{\otimes N})/N$ is shown as a function of the number of qubits $N$. The top panel shows the internal coherence per qubit at different inverse temperatures $\beta$ for the maximally coherent state $\ket{\gamma_S}\bra{\gamma_S}^{\otimes N}$, where $\ket{\gamma_S}=(\ket{0} + e^{-\beta\omega_0/2} \ket{1})/\sqrt{Z_S}$. The lower panel shows the internal coherence per qubit given the parameter values that yield the optimal efficiency $\eta_C$ in our protocol for each system size. In both cases, the internal coherence per qubit is maximized for $N=4$ qubits. While the internal coherence per qubit is maximized for $\beta=0$ for the maximally coherent state, the optimal efficiency is obtained for intermediate temperatures.}
    \label{fig:intcoh per qubit}
\end{figure}

\section{Thermodynamic analysis of the coherence-driven heat engine}
\label{section:Thermodynamic analysis of the coherence-driven heat engine}
\subsection{Coherence conservation}
Based on Ref.~\cite{Latune2020}, we decompose the system's von Neumann entropy change rate $S(\rho_S(t))=-\Tr[\rho_S(t)\ln{\rho_S(t)}]$ into the entropy production rate $\Pi_S=-\Tr[\dot{\rho}_S\left(\ln{\rho_S}-\ln{\gamma_S}\right)]$ and heat exchange rate $\Phi_S=-\Tr[\dot{\rho}_S\ln{\gamma_S}] =-\Tr[\dot{\rho}_S (-\beta H_S - \ln Z_S)] = \beta \Tr[\dot{\rho}_S H_S]$ for a time-independent system Hamiltonian $H_S$.

For our case of a purely coherence-driven engine, the classical energy distribution is equal to the Gibbs state at all times, i.e., $\Delta(\rho_S(t)) =\gamma_S$ and $\Delta(\dot{\rho}_S(t)) = 0$. This leads to zero heat flow $\Phi_S = \beta \Tr[\dot{\rho}_S(t) H_S] = \beta \Tr[\Delta(\dot{\rho}_S(t)) H_S] =0$ as the heat exchange rate only depends on the change in the diagonal elements. We also note that the entropy production rate can be expressed in terms of the coherence change rate
\begin{equation}
    \Pi_S(t)=-\dot{C}_{\rm tot}(\rho_S(t)),
\end{equation}
under the condition $\Delta(\rho_S(t)) =\gamma_S$. This can be derived from the expression $C_{\rm tot}(\rho_S(t)) = S(\rho_S(t) \| \Delta(\rho_S(t))) = S(\rho_S(t) \| \gamma_S) = {\rm Tr}[{\rho}_S (\ln\rho_S - \ln\gamma_S)]$, which leads to $\dot{C}_{\rm tot}(\rho_S(t)) = \frac{d}{dt} {\rm Tr}[{\rho}_S (\ln\rho_S - \ln\gamma_S)] = {\rm Tr}[\dot{\rho}_S (\ln\rho_S - \ln\gamma_S)] = -\Pi_S$ by noting that ${\rm Tr} [\rho_S(t) \frac{d}{dt}\ln\rho_S(t)] = {\rm Tr} [\dot{\rho}_S(t)]=0$ and ${\rm Tr}[\rho_S \frac{d}{dt}\ln\gamma_S] = 0$.

Furthermore, various coherence conservation laws for \textit{athermal operations} were studied in Ref.~\cite{Latune2020} by considering the entropy production of the full system. These athermal operations are defined via initial separability $\rho_{SB}(0)=\rho_S(0)\otimes\rho_B$, energy conservation $[U,H_S+H_B]=0$ and stationarity of the initial bath state $[H_B,\rho_B]=0$. Accordingly, in our external coherence charging protocol, we also arrive at the conservation law in terms of the changes in external coherence of the system (bath) $\Delta C_{\rm ext}(\rho_{S(B)})$
\begin{equation}
    -\Delta  C_{\rm ext}(\rho_{S})-\Delta  C_{\rm ext}(\rho_{B})=C_{\rm ext}(\rho_{S:B}),
\label{eq:coh_conserv}
\end{equation}
which becomes the \textit{correlated external coherence} of the system and bath $C_{\rm ext}(\rho_{S:B})$ studied in Ref.~\cite{Tan2016}.

\subsection{Efficiency}
\begin{figure}[b]
    \centering
    \includegraphics[scale=0.55]{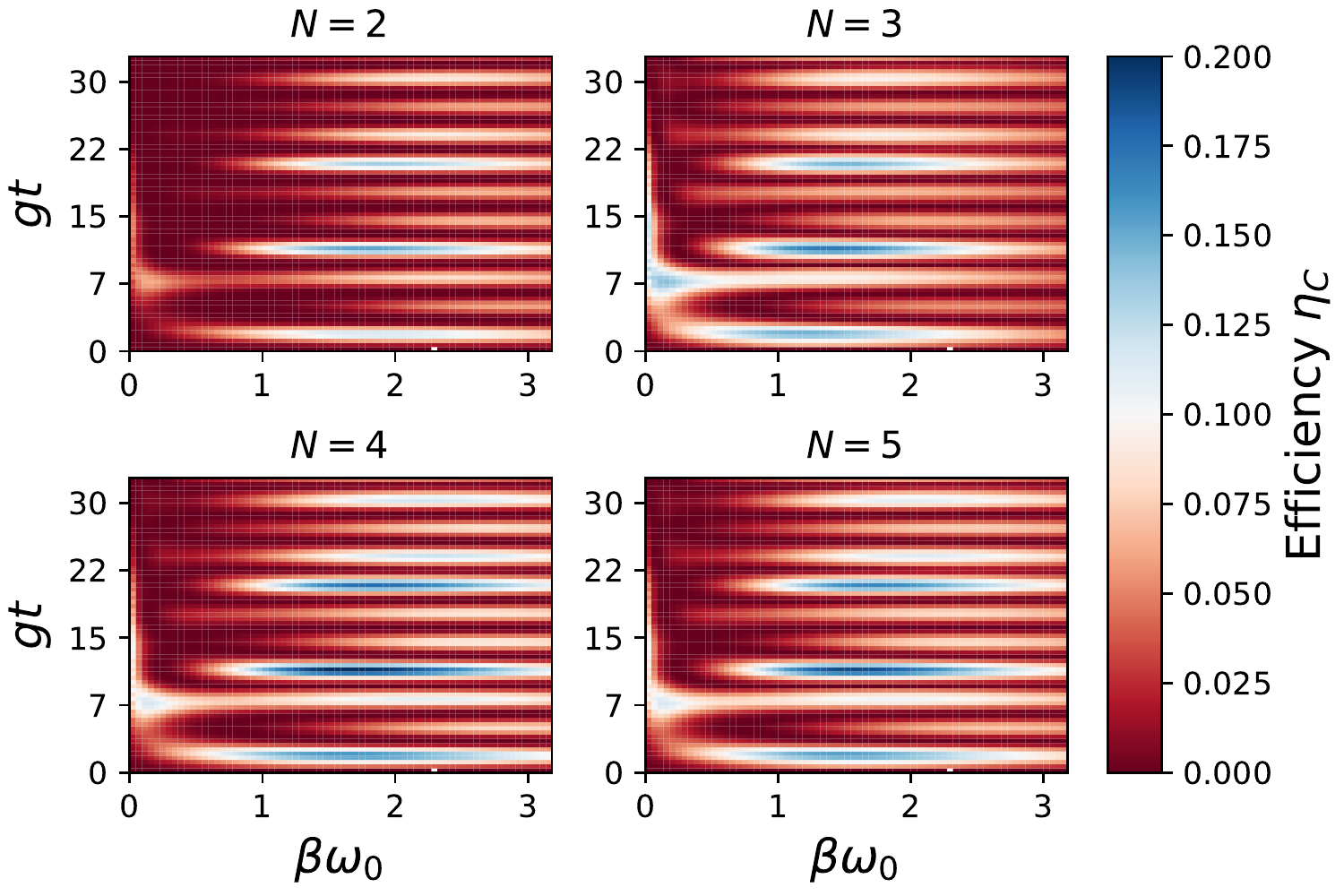}
    \caption{The efficiency of the single-copy coherence charging protocol $\eta_C$ is shown as a function of inverse temperature $\beta\omega_0$ and coupling strength $gt$ for various $N$. It is shown that efficiency is maximized for $N=4$ qubits at intermediate temperatures.}
    \label{fig:efficiency plots}
\end{figure}
\begin{figure}[b]
    \centering
    \includegraphics[scale=0.5]{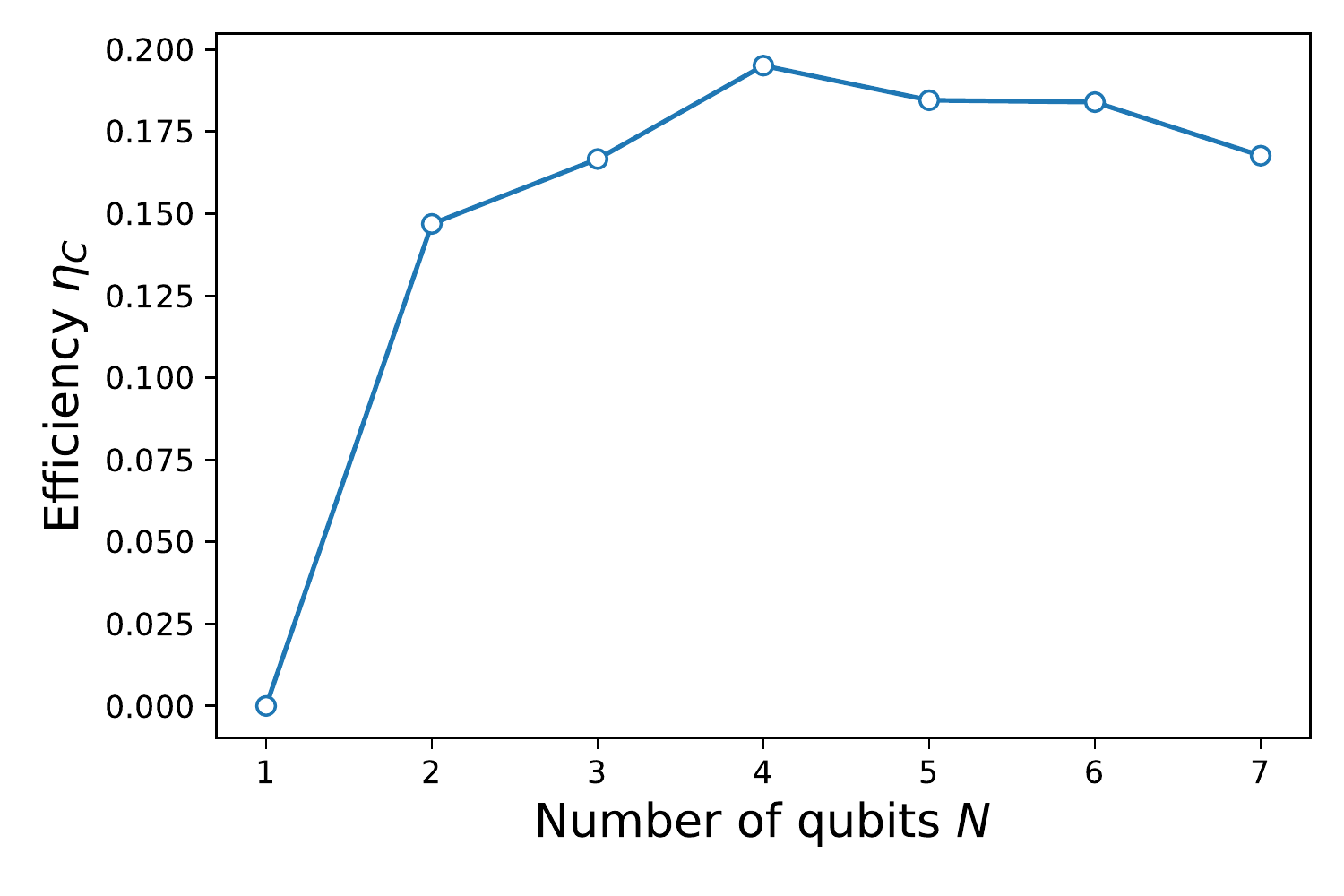}
    \caption{The efficiency $\eta_C$ of the single-copy coherence charging protocol is shown as a function of the number of qubits $N$. The inverse temperatures $\beta\omega_0$ and coupling strengths $gt$ were chosen such that $\eta_C$ is optimized for each $N$. The maximum efficiency is reached for $N=4$ qubits.}
    \label{fig:indiv_charging eta vs N}
\end{figure}
To quantify the performance of the engine, we define the efficiency analogous to classical heat engines by replacing heat flow with coherence flow. The fact that we can do this is corroborated by the conservation law of external coherence in Eq.~\eqref{eq:coh_conserv} and the absence of heat flow, which decouples the notions of \textit{both} internal and external coherence flows from heat flows~\cite{Latune2020}. More precisely, the \textit{efficiency} of the coherence-driven engine $\eta_C$ is expressed as
\begin{equation}
    \eta_C=\frac{W_{\rm coh}}{Q_{\rm coh}^{\rm (in)}},
\end{equation}
where $W_{\rm coh}$ is the extracted coherent work and $Q_{\rm coh}^{\rm (in)}$ is the input coherence flow from the coherence baths. Explicitly, the coherence flow in the unit of energy is given as  $Q_{\rm coh}^{\rm (in)} = -k_B T \Delta C_{\rm ext}(\rho_B)$, in terms of the coherence change of the bath state multiplied by the temperature $k_B T = 1/\beta$. To reiterate, the minus sign here accounts for the coherence flow \textit{into} the system. If each qubit is charged with coherence identically, then for $N$ qubits using $N$ different baths we obtain: 
\begin{equation}
    Q_{\rm coh}^{\rm (in)}=\sum_{k=1}^{N}Q^{\rm (in)}_{{\rm coh}, k} = - N k_B T \Delta C_{\rm ext}(\rho_B(t)),
    \label{eq:coherentflow}
\end{equation}
where $\rho_B(t)$ is the bath state after each qubit's optimal charging protocol. Using Eqs.~\eqref{eq:extcoh}, \eqref{eq:coherentwork} and \eqref{eq:coherentflow} we arrive at the efficiency
\begin{equation}
    \eta_C= \frac{W_{\rm coh}}{Q_{\rm coh}^{\rm (in)}}=-\frac{C_{\rm int}(\rho_S^{(N)})}{N\Delta C_{\rm ext}(\rho_B)}=\frac{C_{\rm int}(\rho_S^{\otimes N})}{N S(\rho_B(t))},
\end{equation}
where one can explicitly show that $0 \leq \eta_C \leq 1$ (see Appendix~\ref{appendix:d}).
%\hk{How to state that $\eta_C\leq 1$? Shall we include the proof somewhere? Or list the relevant equations invoked?}
 Thus, operating the cycle most efficiently does not only involve optimizing the internal coherence but also taking into account the coherence flow from the bath. Numerical optimisation of the efficiency shows that small coupling strengths and intermediate temperatures are favored for the protocol. 
 As demonstrated in Fig.~\ref{fig:intcoh per qubit}, the internal coherence per qubit for the coherent state $C_{\rm int}(\ket{\gamma_S}\bra{\gamma_S}^{\otimes N})/N$ is maximized for $N=4$ qubits. Similarly, the same result is obtained under the optimization of efficiency. Large coupling strengths and low-temperature result in a low efficiency, which allows us to narrow the optimization range to $gt\in [0,30],\,\,\,\beta\omega_0\in [0,3]$, as shown in Fig.~ \ref{fig:efficiency plots}. While the optimal efficiency is obtained for different $(\beta\omega_0,gt)$ parameter values for a given system size of $N$ qubits, $N=4$ is also shown to be the most efficient system size with $(\beta\omega_0,gt)=(1.57,11.22)$. 

Crucially, coherence is not flowing entirely into the system as external coherence but becomes locked in the system-bath correlations. Since the amount of system-bath correlations grows with the duration of the protocol, shorter protocols can avoid the loss of coherence inaccessible to work and are hence more efficient. In addition, higher temperatures increase the coherence change of the bath (see Fig.~\ref{fig:Coh_charge}(b)), thus lowering the temperature for optimal efficiency as opposed to optimal coherence charging. Qualitatively, higher bath temperatures, effectively corresponding to a higher-dimensional bath, contribute to coherence loss in the system-bath correlations. After the work extraction, the excessive coherence in the system-bath correlations is discarded and transferred as output coherence flow $Q_{\rm coh}^{\rm (out)}$ into a thermal bath specified by the Gibbs state $\gamma_B$. 

The optimized efficiency $\eta_C$ is plotted as a function of qubit number $N$ in Fig.~\ref{fig:indiv_charging eta vs N} featuring a maximum value for a small and controllable system size of four qubits, which might be advantageous for experimentalists. 

\section{Conclusion}
We investigated a thermodynamic cycle, from which work can be extracted solely from quantum coherence. As the proposed protocol does not induce heat flow from the coherence bath while charging the coherence of the system, such a heat engine can be regarded as purely quantum, which does not have any counterpart in classical thermodynamics. For the work extraction process, only the contribution from internal coherence can be used. As a particular coherence charging mechanism, a chain of qubits coupled to a coherent bath, as described by the Jaynes-Cummings model, is optimally charged using single-copy coherence charging. Both temperature and coupling strength dependence on the charging performance and efficiency were investigated.

We found that the system state's coherence is maximally charged when the system interacts with the coherence bath storing an intermediate degree of coherence. We also note that the internal coherence is maximally activated from the $N$ copies of qubit states when $N=4$. The protocol parameters yielding the optimal efficiency do not coincide with those obtained for optimal coherence charging, with lower temperatures necessary to maximize coherence flow into the system and minimize coherence flow into the system-bath correlations. 

While in our study no collective charging protocol using the Tavis-Cummings model was found to enhance the efficiency of the cycle, a thorough investigation into a more complex model of the coherence bath and its interaction~\cite{Mayo_2022} with the system would be an interesting pathway for further research. One simple extension might be an investigation of the effect of purely coherence-driven engines coupled to baths with a degenerate energy spectrum~\cite{Latune2021}. This could allow for the direct charging of internal coherence instead of simply activating it from externally charged copies.

\acknowledgements
H.K. is supported by the KIAS Individual Grant No. CG085301 at Korea Institute for Advanced Study.

\appendix
\section{Proof of Observation~\ref{ob:observation1}}\label{appendix:a}
\begin{proof}
 We take the initial state $\gamma^{\otimes N}_S=Z_S^{-N}\sum_{E_k}e^{-\beta E_k}\ket{E_k}\bra{E_k}$ and the bath state $\rho_B = \sum_{\epsilon_l, \epsilon_{l'}} (\rho_B)_{ll'} \ket{\epsilon_l} \bra{\epsilon_{l'}}$ with $\Delta(\rho_B) = \gamma_B = \sum_{\epsilon_l} \frac{ e^{-\beta \epsilon_l} }{Z_B} \ket{\epsilon_l} \bra{\epsilon_{l}}$, where $\ket{E_k}$ and $\ket{\epsilon_l}$ are the energy eigenstates for the $N$-qubit system and the bath system, respectively. By employing the completeness and orthonormality of the eigenstates we arrive at:
\begin{widetext}
\begin{equation}\label{eq:Appd1}
\begin{aligned}
      \Delta(\rho_S^{(N)}(t))
      &=\sum_{E_i}\ket{E_i}\bra{E_i} \times \bra{E_i} \rho_S(t) \ket{E_i}\\
      &= \sum_{E_i}\ket{E_i}\bra{E_i} \times \bra{E_i}{\rm Tr}_B \left[ U(t) \left( \gamma_S^{\otimes N}\otimes \rho_B \right) U^\dagger(t) \right]\ket{E_i}\\
      &= \sum_{E_i}\sum_{\epsilon_j} \ket{E_i}\bra{E_i} \times \bra{E_i, \epsilon_j} U(t) \left(\gamma_S^{\otimes N}\otimes \rho_B \right) U^\dagger (t) \ket{E_i, \epsilon_j}\\
      &= \sum_{E_i, E_k}\sum_{\epsilon_j,\epsilon_l,\epsilon_{l'}}\frac{e^{-\beta E_k}}{Z_S^N} (\rho_B)_{ll'} \ket{E_i}\bra{E_i} \times \bra{E_i, \epsilon_j} U(t)\ket{E_k,\epsilon_l}\bra{E_k,\epsilon_{l'}}  U^\dagger (t) \ket{E_i, \epsilon_j}\\
      &= \sum_{E_i, E_k}\sum_{\epsilon_j,\epsilon_l,\epsilon_{l'}}\frac{e^{-\beta E_k}}{Z_S^N} (\rho_B)_{ll'} \ket{E_i}\bra{E_i} \times \delta\left(E_i+\epsilon_j -(E_k+\epsilon_l)\right)\delta\left(E_k+\epsilon_{l'}-(E_i+\epsilon_j)\right)\\
      &\qquad\qquad\qquad\qquad \times \bra{E_i, \epsilon_j} U(t)\ket{E_k,\epsilon_l}\bra{E_k,\epsilon_{l'}}  U^\dagger (t) \ket{E_i, \epsilon_j}\\
      &= \sum_{E_i, E_k}\sum_{\epsilon_j, \epsilon_l, \epsilon_{l'}}\frac{e^{-\beta (E_i + \epsilon_j - \epsilon_{l'}})}{Z_S^N} (\rho_B)_{ll'} \ket{E_i}\bra{E_i}
      \times \bra{E_i, \epsilon_j} U(t)\ket{E_k,\epsilon_l} \bra{E_k,\epsilon_{l'}}  U^\dagger (t) \ket{E_i, \epsilon_j} \times \delta(\epsilon_l - \epsilon_{l'})\\
      &= \sum_{E_i, E_k}\sum_{\epsilon_j, \epsilon_l, \epsilon_{l'}}\frac{e^{-\beta (E_i + \epsilon_j - \epsilon_{l'}})}{Z_S^N} \frac{e^{-\beta \epsilon_l}}{Z_B} \ket{E_i}\bra{E_i}
      \times \bra{E_i, \epsilon_j} U(t)\ket{E_k,\epsilon_l} \bra{E_k,\epsilon_{l'}}  U^\dagger (t) \ket{E_i, \epsilon_j} \times \delta(\epsilon_l - \epsilon_{l'})\\
      &= \sum_{E_i}\frac{e^{-\beta E_i}}{Z_S^N}\ket{E_i}\bra{E_i} \times \sum_{\epsilon_j}\frac{e^{-\beta\epsilon_j}}{Z_B} \times \sum_{E_k, \epsilon_l} \bra{E_i, \epsilon_j} U(t)\ket{E_k,\epsilon_l} \bra{E_k,\epsilon_{l}}  U^\dagger (t) \ket{E_i, \epsilon_j}  \\
      &= \sum_{E_i}\frac{e^{-\beta E_i}}{Z_S^N}\ket{E_i}\bra{E_i} \times \sum_{\epsilon_j}\frac{e^{-\beta\epsilon_j}}{Z_B} \\
      &= \gamma_S^{\otimes N},
\end{aligned}
\end{equation}
\end{widetext}
where $\delta(\epsilon_l-\epsilon_{l'})$ arises as there only exists a single energy level $\epsilon_l = \epsilon_{l'} = E_i + \epsilon_j - E_k$ to meet the conditions $E_i+\epsilon_j = E_k+\epsilon_l$ and $E_k+\epsilon_{l'} = E_i+\epsilon_j$ when the energy levels $\{ \epsilon_l \}$ are non-degenerate. We also use that the diagonal components of the initial bath state are given as $(\rho_B)_{ll} = e^{-\beta \epsilon_{l}/Z_B}$ and use the completeness relation $\sum_{E_k, \epsilon_l} \ket{E_k, \epsilon_l} \bra{E_k, \epsilon_l} = \mathbb{1}$ to cancel out $U(t)$ and $U^\dagger(t)$ in the equation.

Hence, we showed $\Delta(\rho_S^{(N)}(t))=\gamma_S^{\otimes N}$. Analogously, it can be shown that $\Delta(\rho_B(t))=\gamma_B$ when $H_S$ is non-degenerate.
\end{proof}

We show that Observation~\ref{ob:observation1} does not hold when the condition for the initial system state $\rho_S(0) = \gamma_S$ is not satisfied. We consider the simplest case where a single qubit system is interacting with a bosonic bath under the Jaynes-Cummings interaction defined in Eq.~\eqref{eq:JC_interaction}. For a generic initial state $\rho_S(0)$ and a coherent bath state $\rho_B$ the final system state at time $t$ becomes
$$
\begin{aligned}
\rho_S(t) &= {\rm Tr}_B [U(t) (\rho_S(0) \otimes \rho_B) U^\dagger(t)]\\
&= e^{-i H_S t} {\rm Tr}_B [U_I(t) (\rho_S(0) \otimes \rho_B) U_I^\dagger(t)] e^{i H_S t},
\end{aligned}
$$
with $H_S = (\omega_0/2) \sigma_z$. By defining $\rho_{ij}(t) = \bra{i} \rho_S(t) \ket{j}$ with $i,j = 0, 1$ and solving the equation of each component using Eq.~\eqref{eq:UI}, we obtain the full information of the final system state as
\begin{equation}
\begin{aligned}
\rho_{00}(t) &= \sum_p \bigg[ \cos^2(gt \sqrt{p}) \rho_{00}(0) \bra{p} \rho_B \ket{p} \\
&~~\quad + \sin^2(gt\sqrt{p}) \rho_{11}(0) \bra{p-1}\rho_B \ket{p-1}\\
&~~\quad +i \sin(gt\sqrt{p})\cos(gt\sqrt{p}) \rho_{01}(0) \bra{p}\rho_B\ket{p-1}\\
&~~\quad -i \sin(gt\sqrt{p})\cos(gt\sqrt{p}) \rho_{10}(0) \bra{p-1}\rho_B\ket{p} \bigg]
\end{aligned}
\label{eq:app_diag}
\end{equation}
and
\begin{equation}
\begin{aligned}
&\rho_{01}(t) \\
&= e^{-i\omega_0 t}\sum_p \bigg[ i \sin(gt {\sqrt{p+1}}) \cos(gt\sqrt{p}) \rho_{00}(0) \bra{p}\rho_B\ket{p+1}\\
&~~\quad -i \sin(gt\sqrt{p})\cos(gt\sqrt{p+1}) \rho_{11}(0) \bra{p-1}\rho_B \ket{p}\\
&~~\quad + \cos(gt\sqrt{p})\cos(gt\sqrt{p+1}) \rho_{01}(0) \bra{p}\rho_B\ket{p}\\
&~~\quad + \sin(gt\sqrt{p})\sin(gt\sqrt{p+1}) \rho_{10}(0) \bra{p-1}\rho_B\ket{p+1} \bigg],
\end{aligned}
\end{equation}
by noting that $\rho_{11}(t) = 1 - \rho_{00}(t)$ and $\rho_{10}(t) = \rho^*_{01}(t)$.

\begin{figure}[t]
    \centering
    \includegraphics[scale=0.5]{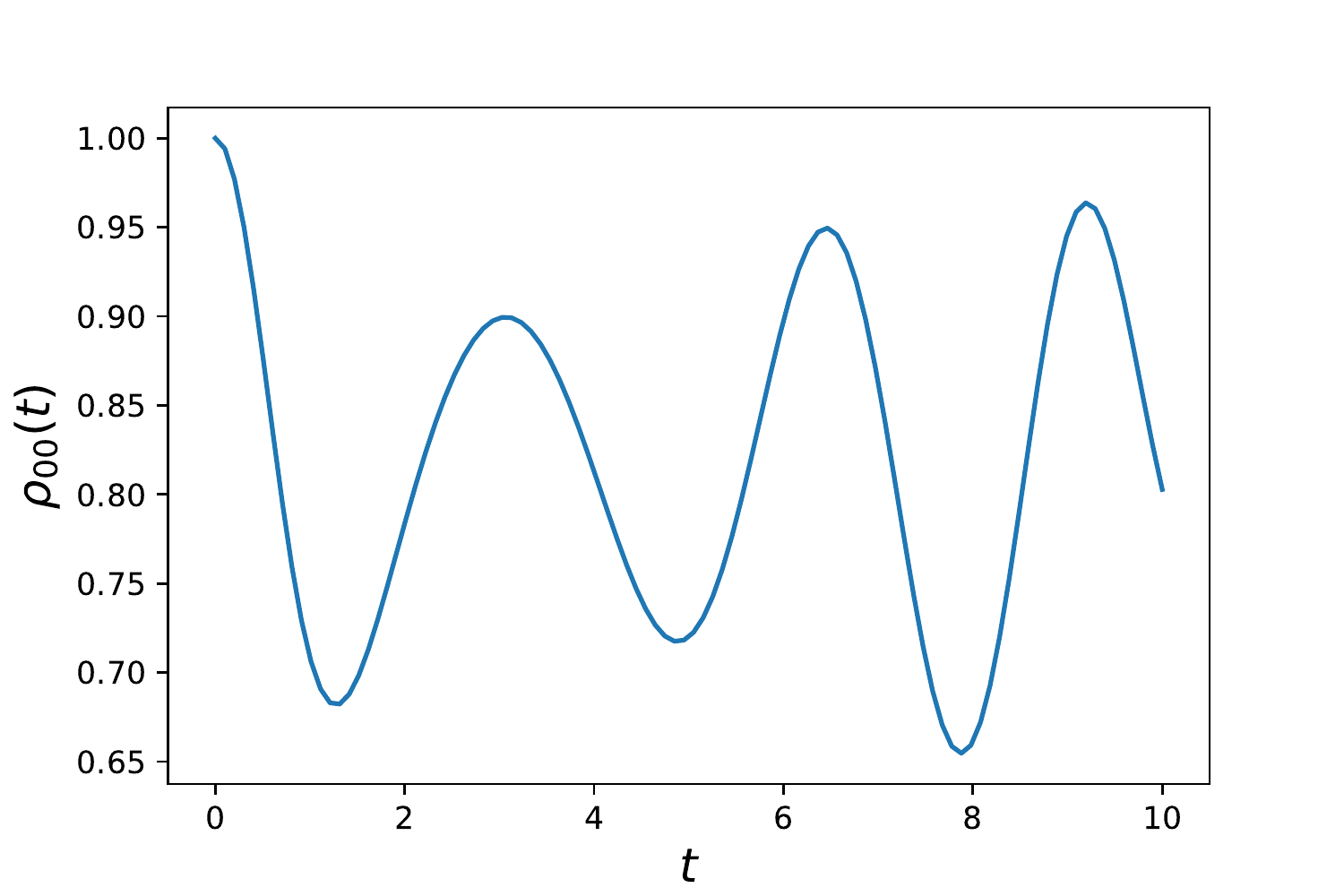}
    \caption{The ground state population of the system $\rho_{00}(t)$ after time $t$ with an initial state $\rho_S(0) = \ket{0}\bra{0}$. The system parameters are taken to be $\beta\omega_0=1$ and $g=1$.}
    \label{fig:app_1}
\end{figure}

We first consider the case $\Delta(\rho_S(0)) \neq \gamma_S$ by taking $\rho_S(0) = \ket{0}\bra{0}$ and $\rho_B = \ket{\gamma_B}\bra{\gamma_B}$. A straightforward calculation from Eq.~\eqref{eq:app_diag} using $\langle p |\gamma_B \rangle = e^{-(\beta \omega_0 /2)p}/\sqrt{Z_B}$ leads to 
$$
\rho_{00}(t) = (1/Z_B) \sum_p e^{- (\beta \omega_0)p } \cos^2(gt\sqrt{p}),
$$
which is in general not the same as the initial population $\rho_{00}(0) = 1$ (see Fig.~\ref{fig:app_1}). This implies that one cannot keep the diagonal elements invariant for an initial state $\Delta(\rho_S(0)) \neq \gamma_S$.

\begin{figure}[t]
    \centering
    \includegraphics[scale=0.5]{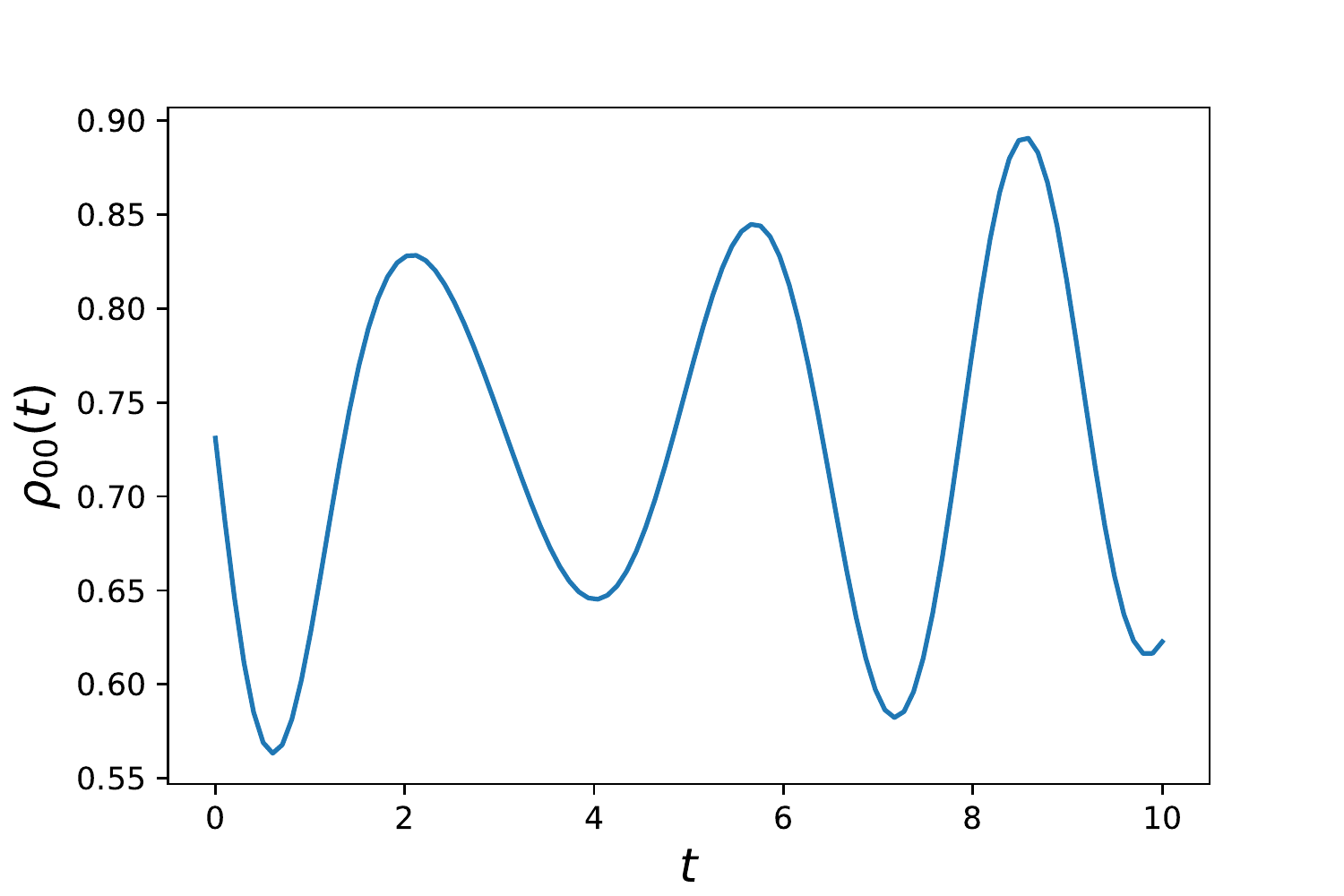}
    \caption{ The ground state population of the system $\rho_{00}(t)$ after time $t$ with an initial state $\rho_S(0) = \gamma_S + i \kappa (\ket{0}\bra{1} - \ket{1}\bra{0})$. The system parameters are taken to be $\kappa=0.3$, $\beta\omega_0=1$, and $g=1$.}
    \label{fig:app_2}
\end{figure}

Second, we consider the case where the initial state already has some coherence, i.e. $\Delta(\rho_S(0)) = \gamma_S \neq \rho_S(0)$, we take $\rho_S(0) = \gamma_S + i \kappa (\ket{0}\bra{1} - \ket{1}\bra{0})$ and $\rho_B = \ket{\gamma_B}\bra{\gamma_B}$. By noting that $\rho_{00}(0) = 1/Z_S$ and $\rho_{11}(0) = e^{-\beta \omega_0}\rho_{00}(0) = e^{-\beta \omega_0}/Z_S$ from the condition $\Delta(\rho_S(0)) = \gamma_S \neq \rho_S(0)$, we obtain 
$$
\rho_{00}(t) = \frac{1}{Z_S} - \frac{\kappa e^{\beta \omega_0/2}}{Z_B} \sum_p e^{-(\beta \omega_0) p} \sin(2gt\sqrt{p}).
$$
As the second term does not vanish, in general, we observe that the ground state population does not remain the same, i.e., $\rho_{00}(t) \neq \rho_{00}(0) =  1/Z_S$ (see Fig.~\ref{fig:app_2}).

\section{Proof of Observation~\ref{ob:observation2}}\label{appendix:b}
\begin{proof}
Let us define the multi-qubit state after interacting with a single coherence bath state $\rho_B = \sum_{\epsilon_l,\epsilon_{l'}} (\rho_B)_{ll'} \ket{\epsilon_l} \bra{\epsilon_l'}$ satisfying $\Delta(\rho_B) = \gamma_B$ (or equivalently, $(\rho_B)_{ll} = e^{-\beta \epsilon_l}/Z_B$) for some time $t$ as
$$
\rho_S^{(N)}(t) = {\rm Tr}_B \left[ U(t) \left(\gamma_S^{\otimes N} \otimes \rho_B \right)U^\dagger(t)\right].
$$
We then take two degenerate energy eigenstates of the $N$-qubit system, $\ket{E_\mu}$ and $\ket{E_\nu}$ in the same energy subspace with energy $E$, such that $H_S^{(N)} \ket{E_\mu} = E \ket{E_\mu}$ and $H_S^{(N)} \ket{E_\nu} =E\ket{E_\nu}$, where $H_S^{(N)} = (\omega_0/2)\, \sum_{k=1}^{N}\sigma_{z|k}$ is the Hamiltonian of the non-interacting $N$-qubit system. We then calculate the overlap between these two eigenstates for the final state as in Eq.~\eqref{eq:Appd1}:
\begin{widetext}
$$
\begin{aligned}
&\bra{E_\mu} \rho_S^{(N)}(t) \ket{E_\nu} \\
      &= \sum_{E_{k_1}, \cdots E_{k_N}}\sum_{\epsilon_j,\epsilon_l,\epsilon_{l'}}\frac{e^{-\beta (E_{k_1} + \cdots + E_{k_N})}}{Z_S^N} (\rho_B)_{ll'} \bra{E_\mu, \epsilon_j} U(t)\ket{E_{k_1}, \cdots E_{k_N},\epsilon_l}\bra{E_{k_1}, \cdots E_{k_N},\epsilon_{l'}}  U^\dagger (t) \ket{E_\nu, \epsilon_j}\\
      &= \sum_{E_{k_1}, \cdots E_{k_N}}\sum_{\epsilon_j,\epsilon_l,\epsilon_{l'}}\frac{e^{-\beta (E_{k_1} + \cdots + E_{k_N})}}{Z_S^N} (\rho_B)_{ll'} \bra{E_\mu, \epsilon_j} U(t)\ket{E_{k_1}, \cdots E_{k_N},\epsilon_l}\bra{E_{k_1}, \cdots E_{k_N},\epsilon_{l'}}  U^\dagger (t) \ket{E_\nu, \epsilon_j}\\
      &\qquad\qquad\qquad\qquad \times \delta(E + \epsilon_j - (E_{k_1} + \cdots E_{k_N} + \epsilon_l) ) \delta(E + \epsilon_j - (E_{k_1} + \cdots E_{k_N} + \epsilon_{l'}) )\\
      &= \sum_{E_{k_1}, \cdots E_{k_N}}\sum_{\epsilon_j,\epsilon_l,\epsilon_{l'}}\frac{e^{-\beta (E+\epsilon_j - \epsilon_l) }}{Z_S^N} (\rho_B)_{ll'} \bra{E_\mu, \epsilon_j} U(t)\ket{E_{k_1}, \cdots E_{k_N},\epsilon_l}\bra{E_{k_1}, \cdots E_{k_N},\epsilon_{l'}}  U^\dagger (t) \ket{E_\nu, \epsilon_j}\times  \delta(\epsilon_l - \epsilon_{l'}) \\
      &= \sum_{E_{k_1}, \cdots E_{k_N}}\sum_{\epsilon_j,\epsilon_l}\frac{e^{-\beta (E+\epsilon_j - \epsilon_l) }}{Z_S^N}\frac{e^{-\beta \epsilon_l}}{Z_B} \bra{E_\mu, \epsilon_j} U(t)\ket{E_{k_1}, \cdots E_{k_N},\epsilon_l}\bra{E_{k_1}, \cdots E_{k_N},\epsilon_{l}}  U^\dagger (t) \ket{E_\nu, \epsilon_j}\\
      &= \sum_{\epsilon_j}\frac{e^{-\beta E }}{Z_S^N}\frac{e^{-\beta \epsilon_j}}{Z_B} \bra{E_\mu, \epsilon_j} U(t)  U^\dagger (t) \ket{E_\nu, \epsilon_j}\\
      &= \frac{e^{-\beta E }}{Z_S^N} \langle E_\mu | E_\nu \rangle \\
      &= \frac{e^{-\beta E }}{Z_S^N} \delta_{\mu, \nu},
\end{aligned}
$$
\end{widetext}
which implies that there is no overlap between any degenerate energy states, i.e.,
$$
P_E \rho^{(N)}_S(t) P_E = \frac{e^{-\beta E }}{Z_S^N} P_E.
$$
As $P_E:=\sum_{H_S^{(N)} \ket{E_\mu} = E\ket{E_\mu}} \ket{E_\mu}\bra{E_\mu}$ is always diagonal in any basis decomposition of $\ket{E_\mu}$, we conclude that no internal coherence can be charged via interaction between the $N$-qubit system and a single coherence bath state $\rho_B$ such that $\Delta(\rho_B)=\gamma_B$.
\end{proof}

\section{More details on the effective finite-size bath approximation}\label{appendix:c}
 While a simple analytical answer may not be found, the effective bath dimension $d^*$ can always be found numerically. In the main part of the paper for all numerical investigations, we assume that $d^*$ was found as appropriate.

Eventually, an analytical expression for the effective bath dimension $d^*$ for the coherence charging of individual qubits can be derived by specifying the desired accuracy $\text{acc}$ of the approximation:
\begin{equation}
    d^*(\beta\omega_0)= \lceil{-\left[1+\frac{1}{\beta\omega_0}\ln{\left(\text{acc}\cdot(e^{-\beta\frac{\omega_0}{2}}+e^{\beta\frac{\omega_0}{2}})\right)}\right]\rceil},
    \label{eq:bathdimension_dstar for single-copy charging}
\end{equation}
\begin{proof}
Dropping the imaginary unit, the coherence elements in Eq.~\eqref{eq:single_rho01} equate to $\delta/Z_S=\mathcal{F}_{\infty}=\lim_{d\to \infty} \mathcal{F}_d$, where $\mathcal{F}_{d}=\sum_{p=0}^{d}e^{-\beta\omega_0 p} \mathcal{F}(\beta\omega_0, gt,p)$ with $\mathcal{F}(\beta\omega_0, gt,p)=(Z_S Z^{d}_B)^{-1}\sum_{p=0}^{\infty} e^{-\beta\omega_0 p} \sin{\left(gt(\sqrt{p+1}-\sqrt{p})\right)}$. We want to find the smallest $d$ such that $\mathcal{F}_{\infty}\approx \mathcal{F}_{d}$. Then $d=d^*$, where $d^*$ is the effective finite bath dimension. This means there exists an $\epsilon_{d^{*}}(\beta\omega_0,gt) >0$ and $d^*(\beta\omega_0, gt) \in \mathbb{Z_+}$ such that $|\mathcal{F}_{\infty}-\mathcal{F}_d| < \epsilon_{d^{*}}(\beta\omega_0, gt) \qquad \forall d\geq d^*(\beta\omega_0)$.
Let us define $c_p:= e^{-\beta\omega_0 p} \mathcal{F}(\beta\omega_0, gt,p)$ so that we can write $\mathcal{F}_{d}= \sum_{p=0}^{\infty}c_p$. We then have $\lim_{n\to \infty}\mathcal{F}_n=\mathcal{F}_{\infty}$, thus we can write $\mathcal{F}_{\infty}=\mathcal{F}_n+R_n$, where the remainder or error is defined to be $R_n:=\mathcal{F}_{\infty}-\mathcal{F}_n = \sum_{p=n+1}^{\infty}c_p$.
We would like to bound the remainder $R_n$ and ideally find an upper bound. To do this, we use the ratio test. For $r_n = c_{n+1}/c_n$ and $r=\lim_{n\to\infty}{|r_n|}<1$ the series converges.

Let us rewrite $R_n$:
\begin{equation}
\begin{aligned}
    R_n &=\sum_{p=n+1}^{\infty}c_p\\
    % = c_{n+1} + c_{n+2}+c_{n+3}+c_{n+4}+...\\
    &= c_{n+1}\left(1 + \frac{c_{n+2}}{c_{n+1}} + \frac{c_{n+3}}{c_{n+1}} + \frac{c_{n+4}}{c_{n+1}} + ...\right)\\
    &= c_{n+1}\left(1 + \frac{c_{n+2}}{c_{n+1}} + \frac{c_{n+3}}{c_{n+1}}\frac{c_{n+2}}{c_{n+2}} + \frac{c_{n+4}}{c_{n+1}}\frac{c_{n+2}}{c_{n+2}}\frac{c_{n+3}}{c_{n+3}} + ...\right)\\
    &= c_{n+1}\left(1 + \frac{c_{n+2}}{c_{n+1}} + \frac{c_{n+3}}{c_{n+2}}\frac{c_{n+2}}{c_{n+1}} + \frac{c_{n+4}}{c_{n+3}}\frac{c_{n+3}}{c_{n+2}}\frac{c_{n+2}}{c_{n+1}} + ...\right)\\
    &=c_{n+1}\left(1 + r_{n+1} + r_{n+1}r_{n+2} + r_{n+1}r_{n+2}r_{n+3} + ...\right).
\end{aligned}
\end{equation}
Using the triangle inequality, we have
\begin{equation}
\begin{aligned}
    |R_n| = |c_{n+1}|\cdot|\left(1 + r_{n+1} + r_{n+1}r_{n+2} + r_{n+1}r_{n+2}r_{n+3} + ...\right)| \\
    \leq |c_{n+1}| \left(1 + |r_{n+1}| + |r_{n+1}r_{n+2}| + |r_{n+1}r_{n+2}r_{n+3}| + ...\right).
\end{aligned}
\end{equation}

If $\{|r_n|\}$ is an increasing function (approaching $r$ as $n\to\infty$, which is satisfied here), we obtain
\begin{equation}
\begin{aligned}
    |R_n| \leq |c_{n+1}| \left(1 + |r_{n+1}| + |r_{n+1}r_{n+2}| + ...\right)\\
    \leq |c_{n+1}| \left(1 + r + r^2 +...\right) = |c_{n+1}|\sum_{k=0}^{\infty}r^k.
\end{aligned}
\end{equation}
If $r<1$ we have $\sum_{k=0}^{\infty}r^k = (1-r)^{-1}$, i.e., $|R_n|\leq (1-r)^{-1}|c_{n+1}|:=\epsilon_n(\beta\omega_0, gt)$. 
Then, from our definition $|R_n|=|\mathcal{F}_{\infty}-\mathcal{F}_n|\leq \epsilon_n(\beta\omega_0, gt)$. Clearly, this is related to the effective finite bath dimension $d^*$ defined for all $gt$ as follows: $|C_{\infty}-C_{d^{*}}|\leq \epsilon_{d^*}(\beta\omega_0,gt)$. For $d\geq d^*$, we also have $|C_{\infty}-C_{d}|\leq \epsilon_{d^*}(\beta\omega_0, gt)$. By specifying the form of $\mathcal{F}$, we can infer the bound $\epsilon_{d^*}(\beta\omega_0, gt)$. For a strict time-independent bound, we can analytically solve this to find $d^*(\beta\omega_0)$ by setting $\text{acc}=\epsilon_{d^*}$ to a fixed number, which serves as the required accuracy for the coherence elements. Finally, we should also take the ceiling (and absolute value) as the dimension is a positive integer.
\end{proof}

\section{Bounds on the engine efficiency}\label{appendix:d}
In this section, we explicitly show that the engine efficiency
$$
\eta_C= \frac{W_{\rm coh}}{Q_{\rm coh}^{\rm (in)}}=-\frac{C_{\rm int}(\rho_S^{(N)})}{N\Delta C_{\rm ext}(\rho_B)}=\frac{C_{\rm int}(\rho_S^{\otimes N})}{N S(\rho_B(t))},
$$
is a valid measure of efficiency in the sense that $0 \leq \eta_C\leq 1$. It is straightforward to observe that $\eta_C \geq 0$ as both $C_{\rm int}(\rho_S^{\otimes N})$ and $S(\rho_B(t))$ are always non-negative.

To prove that $\eta_C \leq 1$, we show that the numerator is upper-bounded as
$$
\begin{aligned}
C_{\text{int}}(\rho_S^{\otimes N})
      &= C_{\text{tot}}(\rho_S^{\otimes N})-C_{\rm ext}(\rho_S^{\otimes N})\\
      &\leq C_{\rm tot}(\rho_S^{\otimes N})\\
      &=S(\rho_S^{\otimes N}||\Delta(\rho_S^{\otimes N}))\\
      &=S(\rho_S^{\otimes N}||\gamma_S^{\otimes N})\\
      &=N S(\rho_S ||\gamma_S)\\
      &=N C_{\rm tot}(\rho_S),
\end{aligned}
$$
where we have used $C_{\rm ext}(\rho_S^{\otimes N})\geq 0$, $\Delta(\rho_S^{\otimes N}) = \gamma_S^{\otimes N}$, and the additivity of the relative entropy under the tensor product. We then obtain
$$
\begin{aligned}
C_{\text{int}}(\rho_S^{\otimes N})
&\leq N C_{\rm tot}(\rho_S) \\
&=N\left(-\Delta C_{\rm ext}(\rho_B)-C_{\rm ext}(\rho_{S:B})\right)\\
&\leq -N\Delta C_{\rm ext}(\rho_B)\\
&=N S(\rho_B)
\end{aligned}
$$
as $C_{\rm ext}(\rho_{S:B}(t))\geq 0$, $C_{\rm tot}(\rho_S(t))=\Delta C_{\rm ext}(\rho_S(t))$, and $\Delta C_{\rm ext}(\rho_B(t))=-S(\rho_B(t))$, which completes the proof that $\eta_C = \frac{C_{\rm int}(\rho_S^{\otimes N})}{N S(\rho_B)} \leq 1$.
\bibliographystyle{apsrev4-1}
\bibliography{bibliography}
\end{document}